\documentclass[]{iopart}      
\usepackage{iopams}
\usepackage{graphicx}
\begin{document}

\title[Search for gravitational wave bursts in LIGO's third science run]
      {Search for gravitational wave bursts in LIGO's third science run}

\author{B~Abbott$^{13}$, R~Abbott$^{13}$, R~Adhikari$^{13}$, J~Agresti$^{13}$, P~Ajith$^{2}$, B~Allen$^{39}$, J~Allen$^{14}$, R~Amin$^{17}$, S~B~Anderson$^{13}$, W~G~Anderson$^{39}$
% ...
\\ (LIGO Scientific Collaboration)}
% ...
\address{$^{2}$ Albert-Einstein-Institut, Max-Planck-Institut f\"ur Gravitationsphysik, D-30167 Hannover, Germany}
% ...
\address{$^{13}$ LIGO - California Institute of Technology, Pasadena, CA  91125, USA}
\address{$^{14}$ LIGO - Massachusetts Institute of Technology, Cambridge, MA 02139, USA}
% ...
\address{$^{17}$ Louisiana State University, Baton Rouge, LA  70803, USA}
% ...
\address{$^{39}$ University of Wisconsin-Milwaukee, Milwaukee, WI  53201, USA}
% ...

\eads{\mailto{lindy@ligo.mit.edu}}

\begin{abstract}
We report on a search for gravitational wave bursts in data from
the three LIGO interferometric detectors during their third science run. The search targets
subsecond bursts in the frequency range 100--1100~Hz for which no waveform
model is assumed, and has a sensitivity in
terms of the {\em root-sum-square} (rss) strain amplitude of $h_\mathrm{rss} \sim 10^{-20} \;
\mathrm{Hz}^{-1/2}$. No gravitational wave signals were detected in the 8 days of analyzed
data.
\end{abstract}


\submitto{\CQG}

\section{Introduction}

Gravitational-wave bursts are generally described as time-varying strain
signals that are of very short duration.
Within the context of LIGO data analysis this term describes
primarily sub-second duration signals with significant power in
the instruments' sensitive frequency band.
Typical sources of this kind of radiation
include astrophysical systems for which the resulting burst waveforms are
either poorly modeled or are completely unknown.  These include the core
collapse of massive stars, the merger phase of binary back hole systems and the
astrophysical engines that power gamma ray bursts.  Other sources of
gravitational-wave bursts exist for which their waveforms are well modeled.
These include black hole ringdowns and bursts resulting from cosmic string
cusps and kinks. 
Gravitational-wave bursts may also result from sources that are completely
unknown or not anticipated.

The Laser Interferometer Gravitational wave Observatory (LIGO) is a network of
interferometric detectors aiming to make direct observations of gravitational
waves~\cite{LIGONIM}.  LIGO is composed of three interferometers at two sites.
Two interferometers, one of 4 km (H1) and another one of 2 km arm length (H2),
are co-located at Hanford, WA.  A third instrument of 4 km arm length (L1) is
located at Livingston, LA.  Each detector is a power-recycled
Michelson interferometer with Fabry-Perot cavities in each of its orthogonal
arms.  These interferometers are sensitive to quadrupolar oscillations in the
space-time metric due to passing gravitational waves.

LIGO commissioning has been interspersed with the
collection of data under stable operating conditions in order to perform
astrophysical gravitational wave searches.  The first science run, called S1,
took place in the summer of 2002 (Aug 23--Sept 9), while two additional runs,
S2 and S3, collected data in 2003 (S2: Feb 14--Apr 14 and S3: Oct 31 2003--Jan
9 2004).  A fourth science run, S4, took place at the beginning of 2005 (Feb
22--Mar 23).  As of May 2005 the instruments are
to within a factor of two of their design expectation in their most
sensitive frequency band.

Three searches for gravitational wave bursts were performed using data
collected by the LIGO instruments in S1 and S2~\cite{s1burst,S2GRB,s2burst}.
These include the first {\em untriggered} search using
35.5 hours of S1 data~\cite{s1burst} and the first {\em triggered} search for
gravitational wave bursts in coincidence with one of the brightest GRB's,
030329, which fortuitously occurred during LIGO's S2 run~\cite{S2GRB}.  In the
most recent publication~\cite{s2burst}, the analysis of 239.5 hours of
data taken while the three LIGO detectors were in simultaneous operation during
S2 was reported.  As in the previous burst searches with the
LIGO detectors, no final candidate events were observed and the search results were
interpreted as an upper limit of 0.26 events per day on the
rate of gravitational wave bursts at the instruments at the 90\% confidence
level.  The all-sky averaged sensitivity of the S2 search for bursts
with significant power in the LIGO sensitivity band (100Hz to 1000Hz)
lies in the range of $h_\mathrm{rss} \sim 
10^{-20} - 10^{-19} \; \mathrm{Hz}^{-1/2}$
root-sum-square (rss) strain amplitude ~\cite{s2burst}.
\begin{figure}[!thb]
\begin{center}
\includegraphics[width=2.5in]{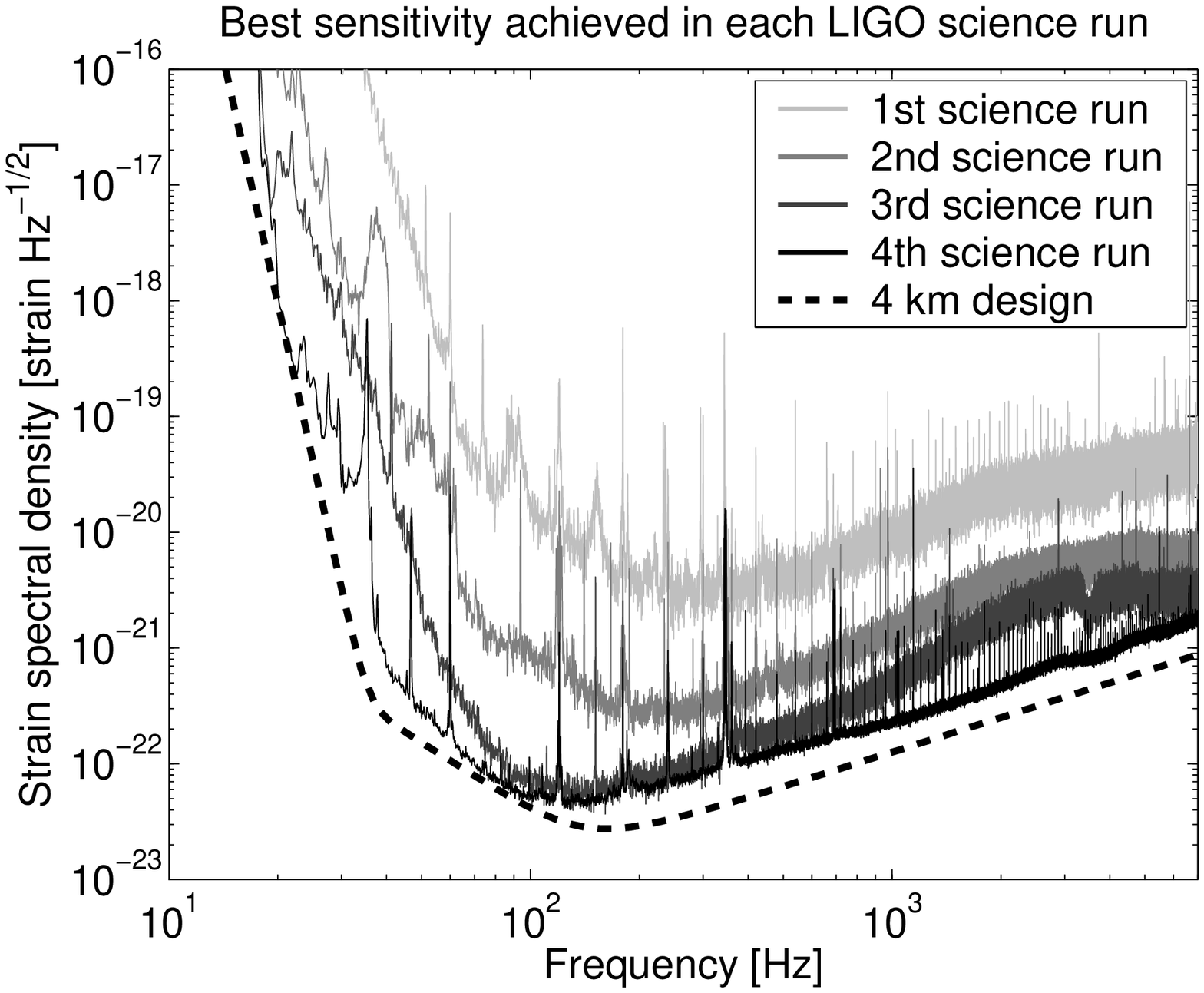}
\includegraphics[width=2.5in]{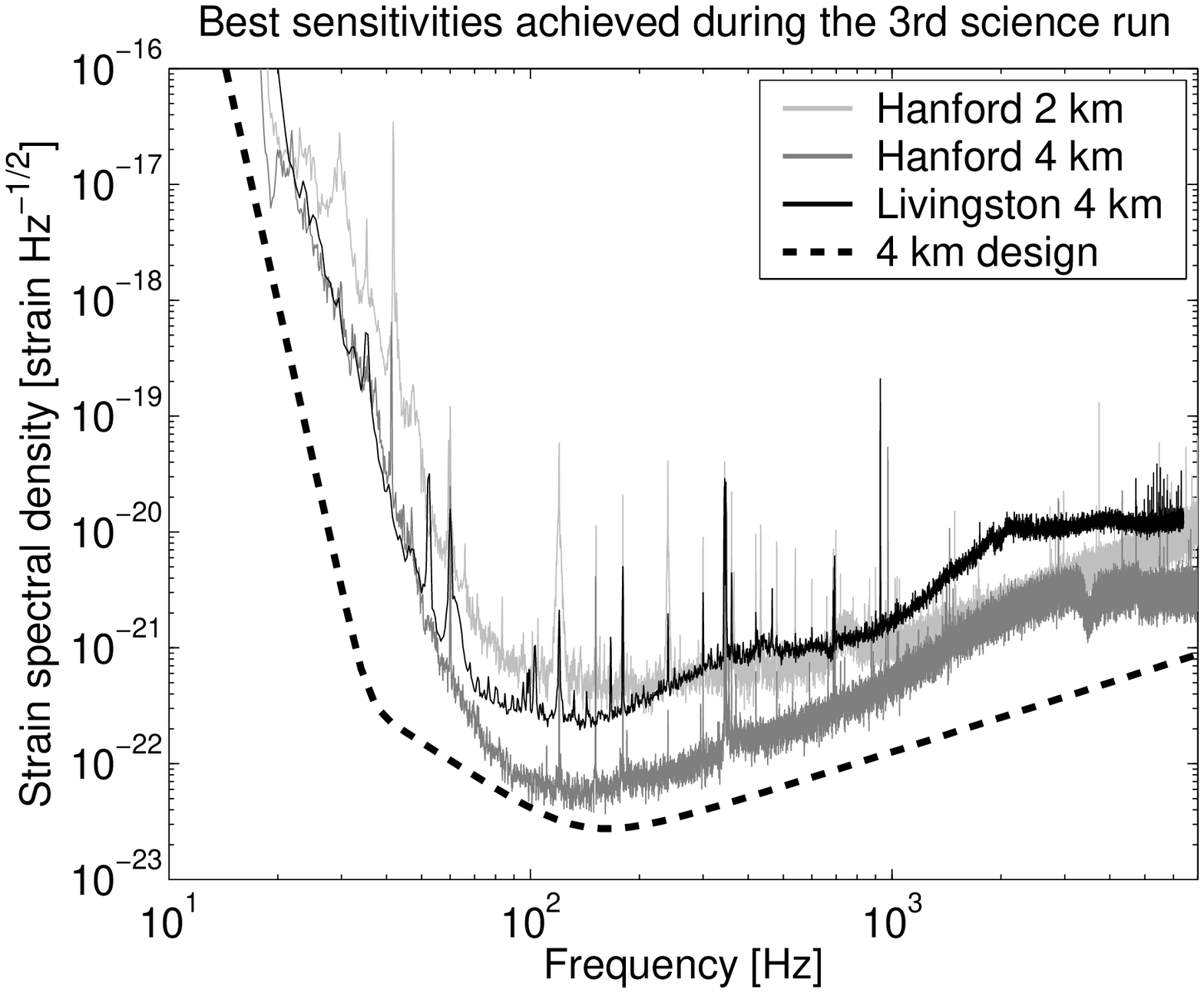}
\caption{Left plot: sensitivity progress of the LIGO 4 km interferometers.  The traces show
the best sensitivity acheived by either of the LIGO interferometers during
each of the four LIGO science runs, along with the 4 km design sensitivity in the LIGO
Science Requirements Document. Right plot: Best sensitivity acheived by each LIGO
interferometer during the third science run.}
\label{fig:noisecurve}
\end{center}
\end{figure}

In this analysis we use data from the S3 run of the LIGO detectors in order
to search for gravitational wave bursts. The S3 run provided data with improved
sensitivity with respect to the previous data taking, as can be seen in figure 1.

\section{Search Pipeline Overview}

The burst search pipeline for the S3 analysis follows closely the procedure
used for the S2 search~\cite{s2burst}. As in S2, the search is restricted to
burst signals that are detectable above the noise in all three
LIGO detectors at once. Therefore, we begin with times when the three
detectors are operating in ``science mode" simultaneously.  This ``triple-coincident"
data set is further reduced by removing periods of data taking when
instrumental artifacts or environmental conditions have been shown to degrade the search.

The Waveburst~\cite{WB1} algorithm is used to identify coincident clusters of
excess power in the wavelet domain across the three gravitational wave data
streams. The triggers generated by Waveburst are checked for amplitude
consistency and then passed to the $r$-statistic~\cite{rstatistic} waveform
consistency test, which uses a normalized cross-correlation statistic to check
for consistent waveform morphology between pairs of detectors.

We estimate the background event rate from accidental noise sources ({\it i.e.},
anything {\em not} directly causing a simultaneous event in the three detectors)
by running the pipeline over
time-shifted data where the gravitational wave data stream from the Livingston
detector is artificially shifted in time with respect to the two Hanford
detectors. It is assumed that the time-shifted noise has similar
characteristics to the unshifted noise, and that the instrumental
behavior is approximately stationary over the range of time shifts (up to two
minutes). To check this assumption, we verify that the distribution of event
counts at the various nonzero time shifts is consistent with a Poisson process.
Detection efficiencies for a variety of ad-hoc and model-based
waveforms are measured by running the pipeline over the real detector data,
with software injections added to the timeseries. The efficiencies measured are
checked against those of physical hardware injections carried out
during the run.

We tune the parameters of the search algorithms with the goal of maximizing detection efficiency over
the simulated events while maintaining a very low false event rate. Unlike the
S2 analysis, time-shifted data over the entire run is used for tuning instead of
a random subset of ``playground'' data set aside purely for such studies. This
procedure avoids removing a valuable fraction of the data from the
analysis result and reduces the chance that the playground data is
unrepresentative of the entire data set. Once the thresholds and parameters of
the search are decided, we run the pipeline over a new set of time shifts to
estimate the background rate, as well as the unshifted data to search for
candidate gravitational wave events.

\section{Data Selection}

There are 265.1 hours of data with all three detectors operating simultaneously
in science mode, giving a triple-coincident duty cycle of 16\% over the S3 run.
From these, 14.0 hours (5.3\%) are removed due to data acquisition problems:
unwritten data, data-acquisition overflows, and timing and synchronization
errors.

A number of additional instrumental issues were discovered during the analysis
and accounted for in the final data selection.  First,
we ignore the 10 seconds just before loss of optical cavity resonance in any of the interferometers,
as such loss is often preceded by a sudden growth in instrument instability.  Also,
periods of excessive levels of dust at any of the output optical tables of the
interferometers are removed from the analysis. Large transients in
the gravitational wave channel were found to occur during large fluctuations in the
light level stored in the arm cavities; such periods are identified and removed. We
implement two event-by-event vetoes that are used to remove single events that
can be identified with observed instrumental artifacts. The first is a veto
applied to all three detectors on events caused by a calibration line
drop-out. The second is a veto for events occurring simultaneously with a large excursion
in the power-recycling servo loop control signal for H2. Details on the selection
and safety of the event-by-event vetoes can be found in the S3 data quality and
veto paper~\cite{s3dataqual}.  In total, these cuts reduce the data set by an
additional 16.8\%.

The presence of a remaining environmental event at the end of the S2 burst
analysis~\cite{s2burst} underscored the need to monitor
environmental disturbances. In the case of the S2 event, strong coherent signals were
acoustically coupled into the co-located H1 and H2 detectors when a
propeller airplane flew overhead. Although the acoustic coupling was reduced
for S3, airplane signals in the gravitational wave channel were still observed during
our investigations. To automate a search for these acoustical disturbances, we identify periods in
many of the microphone channels with large RMS noise in the 62--100~Hz range.
Periods of high acoustic activity are removed from the analysis at both sites.
A similar RMS-based monitor is used on seismic data from the Hanford site to identify periods of
high seismic activity at frequencies with large coupling to the mirrors. These
two environmental cuts further reduce the data set by 1.5\%.

The above data quality cuts remove 62.5 hours from the original 265.1
hours of triple-coincident livetime. The Waveburst algorithm is able to analyze
95\% of the remaining 202.6 hours, with some loss due to data stream segmentation
and boundary effects of the wavelet transform, resulting in an effective
S3 livetime of 192.2 hours for this burst analysis.

\section{Event Generation}

\subsection{Trigger Generation}

The Waveburst algorithm~\cite{WB1}, also used for the S2 analysis~\cite{WB2},
generates triggers on coincident excess power in the wavelet domain across the
raw gravitational-wave data streams. The data first undergo a complete wavelet
packet decomposition, giving for each detector a uniform time-frequency map of the signal
indexed in time by $i$ and in frequency by $j$. Significant tiles in each decomposition are defined 
by the largest 10\% of wavelet coefficients at each
effective frequency. They are assigned a {\em significance} according to their
energy-determined rank within the set of tiles at fixed frequency $j$:
\begin{equation}
\label{eqn:waveburst_significance}
y_{ij} = -\ln\left(R_{ij} / N\right),
\end{equation}
where the rank, $R_{ij}$, is equal to 1 for the most energetic and $N$ for the least
energetic of the selected $N$ tiles. The significant tiles with closely matching
tiles in time and frequency
across the three data streams are determined to be ``in coincidence", and a clustering
routine clusters nearby tiles from the set of coincident tiles for each detector separately.

These single detector clusters are thus built from the triple-coincident
energy in the wavelet domain. Each cluster of $k$ tiles, $C(k)$, is
characterized by its {\em cluster~significance}, $z$, given by
\begin{equation}
\label{eqn:waveburst_clustersignificance}
z = Y \; - \; \ln\left(\sum^{k-1}_{m = 0}{\frac{Y^{m}}{m!}}\right) \quad \mbox{where} \quad Y = \sum_{i,j \in C(k)}{\!\!y_{ij}} \: ,
\end{equation}
which has an exponential distribution regardless of cluster size. The {\em
trigger significance}, $Z_{g}$, is calculated as the geometric average of the
cluster significances for a particular H1/H2/L1 coincident triplet of clusters.
$Z_{g}$ provides a measure of the confidence of each triple-coincident event trigger,
and is used for future thresholding.

The Waveburst implementation used for S3 has two major improvements over the S2
version. For S2, Waveburst operated on just two data streams, meaning that for
triple-coincidence analysis, the final triggers from the three detector pairs
were subject to yet another coincidence stage. For S3, Waveburst is able to
analyze an arbitrary number of data streams at once, allowing a tighter triple-
coincidence stage prior to clustering. Also during the S2 analysis, Waveburst
searched the wavelet time-frequency map at a fixed resolution of 1/128~sec
$\times$ 64~Hz. While this was well tuned for a region of the parameter space
of interest, other regions suffered from poor matching of the wavelet basis to
simulated bursts, particularly at low frequencies where the choice of simulated
bursts included many waveforms longer than 1/128 seconds. For S3, Waveburst operates on
several additional time-frequency resolutions, essentially running a separate
analysis at each resolution and combining the results at the end. This allows
for better matching of the time-frequency tiles to a much larger parameter
space.

\subsection{Amplitude Consistency}

\begin{figure}[!thb]
\begin{center}
\includegraphics[width=2.5in]{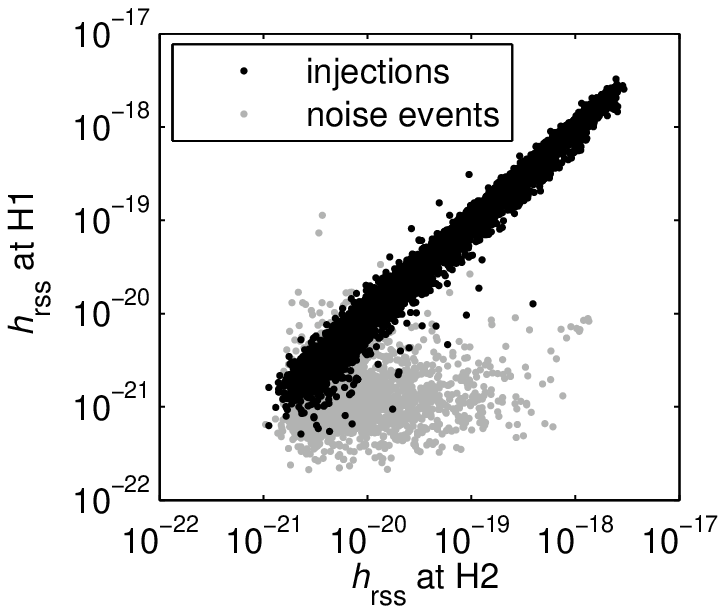}
\includegraphics[width=2.5in]{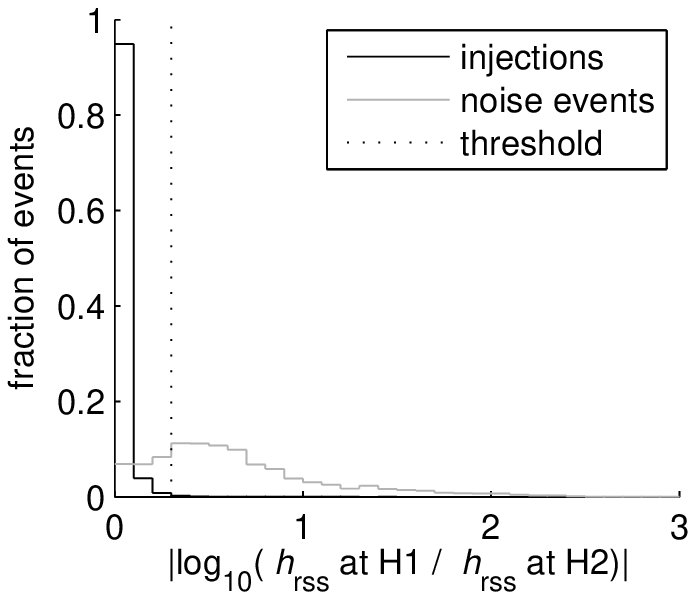}
\end{center}
\caption{Waveburst $h_\mathrm{rss}$ amplitude consistency between H1 and H2 for
injections of simulated signals and for
time-shifted events. On the left is a scatter plot
showing the recorded amplitudes at both detectors for each event. On the
right is a histogram of the absolute value of the logarithm of the ratio of
recorded amplitudes, with a dotted line showing the threshold chosen for an
$h_\mathrm{rss}$ consistency within a factor of two.}
\label{fig:hrssconsistency}
\end{figure}

Because the orientations of the two Hanford interferometers are identical, we
expect to observe the same strain waveform at the two detectors.
Simulations show that the accuracy of signal-energy reconstruction by Waveburst
of a gravitational-wave burst is sufficient to use amplitude consistency to
rule out spurious events.  Based on the performance over simulated signals
shown in figure~\ref{fig:hrssconsistency}, we require the observed
$h_\mathrm{rss}$ amplitudes in the two Hanford detectors to agree within
a factor of two.  This allows us to reject 76\% of the
time-shifted events while maintaining a false rejection rate of just 0.4\%
for simulated bursts.

\subsection{Waveform Consistency}

We use the $r$-statistic test~\cite{rstatistic} to check for waveform
consistency across the three detectors. The test is run over time intervals
triggered by Waveburst as a means of further reducing the background rate. The
test measures the normalized cross-correlation,
\begin{equation}
\label{eqn:rstat}
r = \frac {\sum_i (x_i - \bar{x})(y_{i} - \bar{y})} {\sqrt{\sum_i (x_i-\bar{x})^2}\sqrt{\sum_i (y_{i}-\bar{y})^2}},
\end{equation}
between two whitened gravitational wave strain data time series $\{x_i \}$
and $\{ y_i \}$ with mean values $\bar{x}$ and $\bar{y}$. For uncorrelated white noise
of sufficient length $N_{p}$ such that the central limit theorem applies, we expect the
$r$-statistic values obtained to
follow a normal distribution with zero mean and $\sigma_{p}=1/\sqrt{N_{p}}$. Any
coherent component in the two sequences will cause $r$ to deviate from the
normal distribution.


To compute the $r$-statistic for unknown waveform duration and sky position, we
use integration lengths $N_{p}$ corresponding to 20, 50, and 100 ms, which have been shown to
cover well the burst durations of interest. The integration windows scan over a
region surrounding the Waveburst trigger central time, calculating $r$ using
rectangular windows centered at each time~$j$.  Furthermore, the two data streams may be
shifted by a small amount, $k$, prior to calculating the $r$-statistic. For the H1-H2 pair, $k$
is $\pm$1 ms to account for a small timing error, while for Hanford-Livingston
pairs $k$ takes on values up to $\pm$11 ms to account for all possible physical
light travel times between the sites. For each pair of detectors, the
maximum logarithmic confidence is obtained:

\begin{equation}
\label{eqn:rstat_confidence}
C = \max\left\{-\log_{10}\left[\mathrm{erfc}\left(\left|r^{k}_{pj}\right|\sqrt{\frac{N_{p}}{2}}\right)\right]\right\}
\end{equation}

The parameter $\Gamma$ is then defined as the arithmetic average of
the three values of $C$ from the three detector pairs. This single parameter is used for thresholding to cut events
with low confidence. A final requirement is that the sign of $r$ at maximum
confidence between H1 and H2 must be positive. Otherwise the trigger is discarded since a negative value would
imply opposite phase.
Because L1 is not precisely aligned with the Hanford detectors, it will be
sensitive to different gravitational-wave polarizations and thus different
waveforms.  We therefore do not expect the signals to be 100\% correlated
between the sites.  This is taken into account, in a waveform-dependent way, in
our simulations.

\section{Search Results}

\begin{figure}[!thb]
\begin{center}
\includegraphics[width=2.5in]{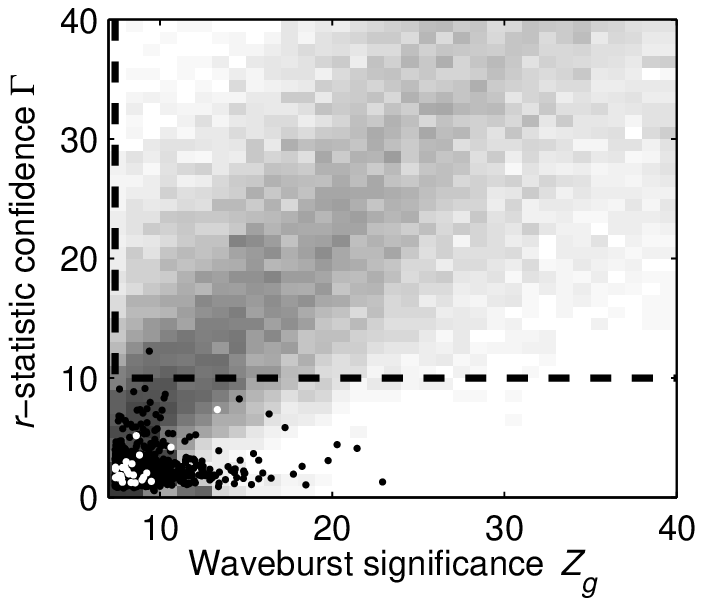}
\includegraphics[width=2.5in]{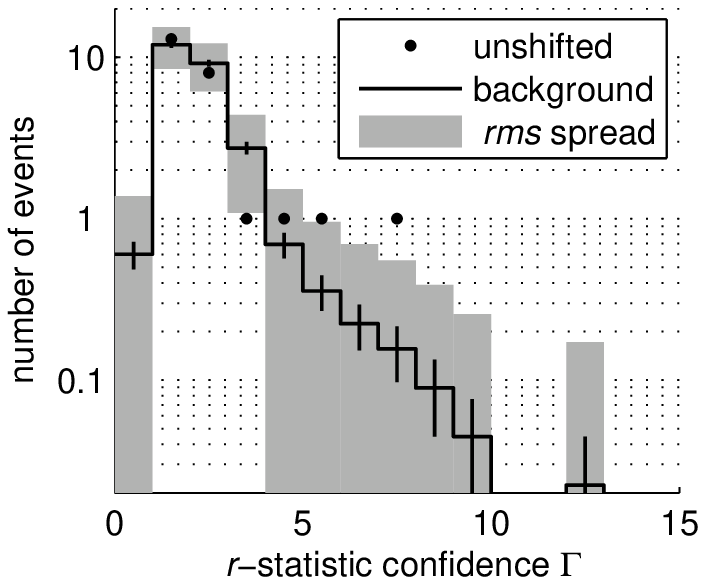}
\end{center}

\caption{Left plot: Measured Waveburst $Z_{g}$ and $r$-statistic $\Gamma$
values for each time-shifted event (black dots) and unshifted event (white dots).
The time-shifted events used to estimate the background of our search are generated over 50
time shifts of the entire S3 data set using 5.0-second steps.  Dotted lines
represent the thresholds on $Z_{g}$ and $\Gamma$ chosen in advance to maintain
a low background event rate while preserving detection efficiency for simulated
events, whose density is represented by the logarithmically weighted 2-D histogram.
In the past, the Waveburst significance has been occasionally shown in
its $\log_{10}$ representation: $Z_{g} / \ln(10)$. Here we follow the
convention used in the S2 paper~\cite{s2burst}.  Right plot: histogram
(circles) of $\Gamma$ values for unshifted events with $Z_g > 7.39$.  The most
significant event has $\Gamma= 7.34$, below our threshold of 10; thus no
events from the analysis at zero time shift remain after all analysis cuts. Stair-step curve: estimated mean
background per bin 
normalized to an observation time equal to that of the unshifted analysis. The
black error bars indicate the statistical uncertainty on the mean background.
The shaded bars represent the expected root-mean-square statistical
fluctuations on the number of unshifted background events in each bin.}

\label{fig:results}
\end{figure}

Preliminary studies over time-shifted S3 data led us to set thresholds on the
Waveburst $Z_{g} \ge$ 7.39, and $r$-statistic $\Gamma \ge$ 10. To estimate the
background rate at these thresholds, we run through the pipeline 50 additional
time shifts of the data using 5-second steps. One time-shifted event survives,
giving an expected background of .02 events per S3 livetime. No
events pass all the analysis cuts in the unshifted data (figure~\ref{fig:results}).

\section{Simulations}

The efficiency of the analysis pipeline is defined as the fraction of
events that would be successfully detected by the pipeline, as a function
of waveform and characteristic amplitude.
Preliminary detection efficiency studies were completed over a randomly
selected 10\% subset of the S3 data. Our simulations include 58 waveforms of
various morphologies: short and long duration sine-Gaussians, Gaussians, cosmic
string cusps~\cite{cusp}, Gaussian windowed band-passed white noise, rising
whistles, black-hole merger simulations~\cite{laz}, and supernova core
collapse simulations~\cite{zm, dfm, ob}. In total, $\sim$100,000 events were
injected over the S3 livetime with durations between 0.1 and 100 milliseconds
and time-frequency area $\Delta t\Delta f$ between 1 and 100, where unity
time-frequency area corresponds to a minimal-uncertainty waveform.

Here we report detection efficiencies of the search pipeline for Gaussian
injections of the form $h(t+t_0)=h_{0} \exp(-t^2/\tau^2)$, with $\tau$ equal to
0.1 ms, and sine-Gaussian injections of the form $h(t+t_0)=h_{0} \sin(2\pi
f_{0} t) \exp(-t^2/\tau^2)$, where $\tau$ is chosen according to
$\tau=$Q$/(\sqrt{2}\pi f_{0})$ with Q=8.9, and $f_{0}$ assumes values of 235,
554, and 849~Hz. These simulated events are generated according to a random,
isotropic sky distribution and have waveforms of purely linear polarization
with random polarization angle.
The strength of the injected events are quantified
by their {\it root-sum-square} (rss) amplitude {\it{at the Earth}} (without
folding in the antenna pattern of a detector) defined by
\begin{equation}\label{eq:hrss}
h_\mathrm{rss} \equiv \sqrt{\int (|h_{+}(t)|^2 + |h_{\times}(t)|^2) \, dt} ~.
\end{equation}
For linearly polarized signals ($h_\times(t)=0$), this is simply the root-sum-square amplitude of
the measured strain for an optimally oriented detector. For a non-optimal
orientation, the measured signal energy is diminished by an antenna factor.

The simulated events are created at constant $h_\mathrm{rss}$, and converted
into detector-specific ADC counts using the known calibration response
function and antenna pattern for each interferometer.  Efficiencies at different 
$h_\mathrm{rss}$ values are evaluated by multiplying the ADC(t) timeseries by
the appropriate factor, adding it to the raw detector data, and running the
combined timeseries through the search pipeline. Table~\ref{tbl:efficiencies}
shows the $h_\mathrm{rss}$ corresponding to 50\% detection efficiency for the
four reported waveforms. We find a factor of $\sim$2 improvement in overall
sensitivity compared to the S2 search.

Alternatively, the efficiency can be evaluated as a function of the
signal energy {\em received} by a given detector, taking the antenna
factor into account.  This can be expressed in terms of the {\em
signal-to-noise ratio} (SNR) that would be measured by an optimal filter,
\begin{equation}
\mathrm{SNR}^2  
=  4 \int_0^{\infty} df \frac{ |F^+ \tilde{h}_+(f) + F^\times \tilde{h}_\times(f)|^2 }{ S(f) } 
\end{equation}
where $\tilde{h}_+(f)$ and $\tilde{h}_\times(f)$ are the {\em two}-sided Fourier
transforms of the two polarization components of the signal,
$F^+$ and $F^\times$ represent the antenna factors,
and $S(f)$ is the {\em one}-sided power spectral density of the noise.
Table~\ref{tbl:efficiencies} shows the SNR in the {\em least} sensitive
detector (calculated event by event using the best noise power spectrum
for each detector during
the run) which yields 50\% detection efficiency.
The majority of the other simulated waveforms
maintain 50\% detection efficiency at SNR {$\sim$5--9} giving us confidence in
the generality of our search pipeline, with respect to match-filtering for known
waveforms.

The systematic uncertainty that results from measuring the efficiency over a
randomly selected 10\% instead of the full data set is not expected to be
large. Furthermore, a higher overlap window (finer increments in time for $j$
and $k$) for the $r$-statistic waveform consistency test was adopted in the analysis of the
full data and not implemented in the efficiency studies, implying that the
efficiencies reported may be underestimated.

\begin{table} [!thb]

\caption{Summary of the S3 pipeline sensitivity to {\it ad hoc} waveforms.
Shown are the 50\% detection efficiencies in terms of
$h_\mathrm{rss}$ [$\mathrm{strain}/\sqrt{\mathrm{Hz}}$]
and in terms of the dimensionless {\em signal-to-noise ratio} (SNR) in the
least sensitive detector.
These values are averages over random sky position and polarization angle.
The equivalent $h_\mathrm{rss}$ values for 50\% detection efficiency at
a comparable expected background event rate for the
same waveforms in the S2 search were 1.5, 2.3, 3.9, and 4.3
$\times$ 10$^{-20}$ $\mathrm{strain}/\sqrt{\mathrm{Hz}}$~\cite{s2burst}.}

\label{tbl:efficiencies}
\begin{indented}\item[]
\begin{tabular}{lcc}
\br
         & \multicolumn{2}{c}{At 50\% detection efficiency:}  \\
Waveform & $h_\mathrm{rss}$ & minimum SNR \\
\mr
sine-Gaussian $f_0$=235~Hz Q=8.9 & 0.9 $\times$ 10$^{-20}$ & 6.0 \\
sine-Gaussian $f_0$=554~Hz Q=8.9 & 1.3 $\times$ 10$^{-20}$ & 5.8 \\
sine-Gaussian $f_0$=849~Hz Q=8.9 & 2.3 $\times$ 10$^{-20}$ & 7.5 \\
Gaussian $\tau$=0.1~ms     &  1.8 $\times$ 10$^{-20}$ & 8.4 \\
\br
\end{tabular}
\end{indented}
\end{table}
\section{Conclusions}

No gravitational wave burst event is observed during the 8 days of LIGO's S3
data that we analyze. Several improvements in the search methodology are
introduced in this analysis. The waveform amplitude consistency test
and the tighter $r$-statistic requirements for H1 and H2 both make
use of the co-location and common orientation of the two Hanford
detectors; information not exploited in the S2 search~\cite{s2burst}. Additionally
the new ability of Waveburst to search at multiple time-frequency resolutions
allows us to maintain sensitivity to a much larger signal space than before. These
improvements are expected to be part of our future burst searches.  The
sensitivity of the search in terms of the {\em root-sum-square} (rss) strain
amplitude is $h_\mathrm{rss} \sim 10^{-20} \; \mathrm{Hz}^{-1/2}$ and reflects the most
sensitive broad-band search for untriggered and unmodeled gravitational wave
bursts to date.

\section*{Acknowledgments}

The authors gratefully acknowledge the support of the United States National
Science Foundation for the construction and operation of the LIGO Laboratory
and the Particle Physics and Astronomy Research Council of the United Kingdom,
the Max-Planck-Society and the State of Niedersachsen/Germany for support of
the construction and operation of the GEO600 detector. The authors also
gratefully acknowledge the support of the research by these agencies and by the
Australian Research Council, the Natural Sciences and Engineering Research
Council of Canada, the Council of Scientific and Industrial Research of India,
the Department of Science and Technology of India, the Spanish Ministerio de
Educacion y Ciencia, the John Simon Guggenheim Foundation, the Leverhulme
Trust, the David and Lucile Packard Foundation, the Research Corporation, and
the Alfred P. Sloan Foundation. This document has been assigned LIGO
Laboratory document number P050043-A-R.

\section*{References}

\end{document}